\documentclass[%
linenumbers,
10,
jpb,
]{iopart}
\usepackage{graphicx}
\usepackage{hyperref}
\usepackage{here}
\usepackage{indentfirst}
\usepackage{dcolumn}
\usepackage{bm}
\usepackage{braket}
\usepackage{hyperref}
\usepackage{txfonts}
\usepackage{ulem}
\usepackage{color}

\newcommand{\ueda}[1]{\textcolor{black}{#1}}
\newcommand{\ota}[1]{\textcolor{black}{#1}}
\newcommand{\yamazaki}[1]{\textcolor{black}{#1}}
\newcommand{\hatada}[1]{\textcolor{black}{#1}}
\newcommand{\hatadarev}[1]{\textcolor{black}{#1}}
\newcommand{\otarev}[1]{\textcolor{black}{#1}}

\begin{document}
\title[Full-potential Multiple Scattering Theory for O 1s PA-MFPADs from CO$^{2+}$]{
Theory of polarization-averaged core-level molecular-frame photoelectron angular distributions: I. A Full-potential method and its application to dissociating carbon monoxide dication}
\author{
F Ota$^1$, 
K Yamazaki$^2$\footnote{Present address: Attosecond Science Research Team, Extreme Photonics Research Group, RIKEN Center for Advanced Photonics, RIKEN, 2-1 Hirosawa, Wako, Saitama, 351-0198, Japan.},
D S\'ebilleau$^3$, 
K Ueda$^4$ and 
K Hatada$^5$}
\address{$^1$ Graduate School of Science and Engineering for Education, University of Toyama, Gofuku 3190, Toyama 930-8555, Japan}
\address{$^2$ Institute for Materials Research, Tohoku University, 2-1-1 Katahira, Aoba-ku, Sendai 980-8577, Japan}
\address{$^3$ D\'epartement Mat\'eriaux Nanosciences, Institut de Physique de Rennes, UMR UR1-CNRS 6251, Universit\'e de Rennes, F-35000 Rennes, France}
\address{$^4$ Institute of Multidisciplinary Research for Advanced Materials, Tohoku University, Katahira 2-1-1, Aoba-ku, Sendai 980-8577, Japan}
\address{$^5$ Faculty of Science, Academic Assembly, University of Toyama, Gofuku 3190, Toyama 930-8555, Japan}
%
%
\ead{hatada@sci.u-toyama.ac.jp}

\date{\today}

\begin{abstract}
We present a theoretical study of the polarization-averaged molecular-frame photoelectron angular distributions (PA-MFPADs) emitted from the $1s$ orbital of oxygen atoms of dissociating dicationic carbon monoxide CO$^{2+}$. 
Due to the polarization-average, the contribution of the direct wave of the photoelectron which represents the largest contribution to the MFPADs is removed, so that the PA-MFPADs clearly show the details of the scattering image of the photoelectron. As a result, it is necessary to employ an accurate theory for the theoretical analysis of the continuum state. In this study, we apply a Full-potential Multiple Scattering theory, where the space is partitioned into Voronoi polyhedra and truncated spheres, in order to take into account the electron charge density outside the physical atomic spheres. We do not use the spherical harmonics expansion of the cell shape functions 
to avoid divergence problems. The potentials in the scattering cells are computed using the Multiconfigurational Second-Order Perturbation Theory Restricted Active Space (RASPT2) method in order to take into account the influence of the core hole in the electron charge density in the final state, so that to a realistic relaxation can be achieved. We show that the Full-potential treatment plays an important role in the PA-MFPADs at a photoelectron kinetic energy of 100 eV. By contrast, the PA-MFPADs are not sensitive to any type of major excited states in the Auger final state.
We also study the dynamics of the CO$^{2+}$ dissociation. We find that the PA-MFPADs dramatically change their shape as a function of the C-O bond length.

\end{abstract}

\submitto{\jpb}
\maketitle
\ioptwocol

\section{Introduction}
\ueda{Core-level photoelectron diffraction (PED) is a well-known technique to probe surface structure~\cite{Woodruff2008}. In order to apply PED to gas-phase molecules, the molecule must be fixed in space. Experimentally, this can be realized by angle-resolved coincidence measurements between core-level photoelectrons and fragment ions~\cite{Shigemasa1995}. With the advent of position sensitive detectors~\cite{Jagutzki2002} and COLTRIMS-Reaction Microscope~\cite{Ullrich2003}, the measure of momentum correlations between core-level photoelectrons and fragment ions to extract the  photoelectron angular distributions in the molecular frame has become a standard approach to study molecular photoionization, see, e.g. the first paper on MFPADs measured by this technique~\cite{Landers2001}.}

\ueda{From a theoretical point of view, it is rather straightforward to calculate MFPADs with the codes available for molecular photoionization studies. For the theoretical tools for molecular photoionization studies in general, see, e.g.,~\cite{Poliakoff2006, Plesiat2012}. One of the widely used approach to calculate MFPADs is to use relaxed-core Hartree-Fock (HF) continuum orbitals in the final-state wavefunction~\cite{Cherepkov2000,Kaiser2020}. The continuum orbitals may be obtained by iteratively solving the Lippmann-Schwinger equation associated with the one electron Schr\"odinger equation with a potential produced by the transition-state orbital~\cite{Saito2002, Rolles2005}. 
To take account of electron correlations, Semenov {\it et al.}~\cite{Semenov2000, Semenov2002} employed random-phase approximations, while \otarev{Lucchese} and collaborators employed multichannel Schwinger configuration interaction (MCSCI) methods~\cite{Stratmann1995, Stratmann1996}. 
Density Functional Theory (DFT) and time-dependent DFT (TDDFT)~\cite{Stenner2005} have also been used to simulate MFPADs~\cite{Plesiat2013, Fukuzawa2019b}. }

\ueda{At higher photoelectron energies, say, $\sim 100$ eV and above, 
 the multiple scattering method within the Muffin-tin approximation, which is a standard tool for PED from  solid surfaces~\cite{Fadley1994},
 has been employed also for calculations of the MFPADs of the gas phase molecules~\cite{Kazama2013}. The multiple scattering method within the Muffin-tin approximation is a powerful tool thanks to the good convergence of the angular momentum  multi-center expansion of the local numerical basis. Therefore, this method allows to treat relatively large polyatomic molecules. We find, however, that it gives a non-negligible error in the MFPADs, especially in the forward scattering directions. This is because the Muffin-tin approximation neglects the covalency and weak potential features out of the sphere which contributes to the small-angle scattering. In the present study, in order to improve the multiple scattering method beyond the Muffin-tin approximation, we introduce a recently-developed Full-potential approach, which fills up the space with Voronoi polyhedra and truncated spherical cells. This allows to take into account the electron charge density outside the physical atomic cells of the target molecule. Note that the cell shape function is not expanded into spherical harmonics to avoid convergence problem.}
 \hatada{This approach has been successfully applied to X-ray Absorption Near Edge Structure (XANES), and has led to substantial improvements for low dimensional systems which have a strong anisotropy~\cite{Hatada2007,Hatada2010, Xu2015}. These improvements over Muffin-tin calculations were observed in particular in the low energy region (about 50 eV from the edge). In the higher energy region, $\gtrsim 100$ eV, there was no noticeable difference observed between the Muffin-tin and the Full-potential XANES calculations.}
 \

\ueda{It should be noted that the MFPADs depend on both the molecular axis and the direction of polarization of the ionizing radiation. Recently, it was noticed that the molecular structural information is captured in the polarization-averaged MFPADs (PA-MFPADs). Due to the polarization-averaging, the contribution of the direct wave of the photoelectron, which is the largest contribution to the MFPADs, is removed so that the PA-MFPADs clearly show the scattering image of the photoelectron. Williams {\it et al.}~\cite{Williams2012} were the first to demonstrate that PA-MFPADs capture the directions of the bonds in polyatomic molecules (CH$_4$). Stimulated by this experimental work, Pl\'esiat {\it et al.} ~\cite{Plesiat2013} calculated the PA-MFPADs for many different molecules in a wide range of photoelectron energies, using DFT. Fukuzawa {\it et al.}~\cite{Fukuzawa2019}, on the other hand, demonstrated that one can extract the bond length of the symmetric linear molecule (CO$_2$) from the PA-MFPADs.}

\ueda{The advent of XFEL~\cite{Emma2010} has stimulated a new surge  of interest in  the study of MFPADs since time-resolved MFPADs or PA-MFPADs using XFEL pulses as ionizing pulses may open a new pathway to make molecular movies. Using a Plane Wave approximation at a electron kinetic energy of 1 keV, Krasniqi {\it et al.}~\cite{Krasniqi2010} theoretically demonstrated that the light-induced structural changes of polyatomic molecules can be extracted from the MFPADs. Kazama {\it et al.}~\cite{Kazama2013}, on the other hand, used the multiple scattering method within the Muffin-tin approximation to demonstrate that it is possible to extract the molecular structure from experimental MFPADs recorded at electron kinetic energies of about 100 eV, thereby opening the way to time-resolved MFPADs measurements with FEL.}
\ueda{Very recently, at the soft X-ray beam line of the European XFEL~\cite{Decking2020}
, the first high-repetition-rate XFEL, Kastirke {\it et al.}~\cite{Kastirke2020} have performed the first successful PA-MFPADs measurements at electron kinetic energies of $\sim 100$ eV, using diatomic molecules O$_2$ as a sample, thereby demonstrating the feasibility of making molecular movies by using a combination of a COLTRIMS-Reaction Microscope and a high repetition rate XFEL.} 

\ueda{ In the present work, we consider the process where the first XFEL pulse produces CO$^{2+}$ via the Auger decay that follows the core ionization of CO and then, the second XFEL pulse kicks out an oxygen $1s$ electron from CO$^{2+}$. In Sections~\ref{sec:Theory} and \ref{sec:Comp}, we derive the expressions of the PA-MFPADs for 
core-level photoemission within the Full-potential multiple scattering theory. The results and their discussion are given in \hatadarev{S}ection~\ref{sec:Results}. We demonstrate then that the PA-MFPADs are not sensitive to the electronic structure, which may differ depending on the Auger final states. Employing a particular electronic configuration of the dicationic state, we show how the PA-MFPADs evolve as a function of the internuclear distance. Finally, we compare the present results to those obtained within the more commonly used Muffin-tin approximation.  This highlights the importance of a proper description of the scattering potentials.} 


\section{Theory}
\label{sec:Theory}
\subsection{Full-potential Multiple Scattering Theory}
\begin{figure}[ht]
\includegraphics[width=\linewidth]{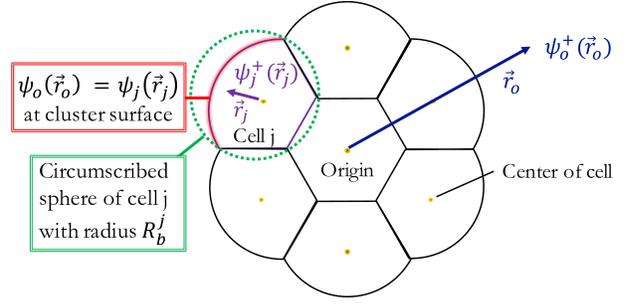}
\caption{ \label{fig:cluster} 
The wavefunction $\psi^{+}_{j} \, ({\bf r}_{j})$ defined in cell $j$ inside the cluster,
the wavefunction $\psi^{+}_{o} \, ({\bf r}_{o})$ for the outer region of the cluster, where we omit $k$.
}
\end{figure}

We start from the identity obtained from the Green's theorem
for the free electron Green's function $G_0^{+}({\bf r}-{\bf r}')$ and
the solution of the Schr\"{o}dinger equation $\psi^{+} ({\bf r};k)$
for the outgoing wave boundary condition at the surface of the cluster \cite{Hatada2010}
\begin{eqnarray}
&&\int_{S_o}
 \Big[
   G_0^+ ({\bf r}_o'-{\bf r}_o;\kappa) \nabla \psi^{+}_o ({\bf r}_o;k) \nonumber \\
  && - \psi^{+}_o ({\bf r}_o;k) \nabla G_0^+ ({\bf r}_o'-{\bf r}_o;\kappa)
 \Big]
 \cdot
 \hat{{\bf n}}_o \,
 d \sigma_{o} 
 \nonumber \\
 &&=
 \sum_j
 \int_{S_j}
 \Big[
   G_0^+ ({\bf r}_j'-{\bf r}_j;\kappa) \, {\nabla} \, \psi^{+}_j ({\bf r}_j;k)
\nonumber \\
&& - \, \psi^{+}_j ({\bf r}_j;k) \, {\nabla} \, G_0^+ ({\bf r}_j'-{\bf r}_j;\kappa)
 \Big]
 \cdot
 \hat{{\bf n}}_j \,
 d \sigma_{j}   
\label{eq:equality}
\end{eqnarray}
where
$\int \hat{{\bf n}}_o \,d \sigma_{o}$ and $\int \hat{{\bf n}}_j \,d \sigma_{j}$ are
integrals over the surface of the cluster and of cell $j$ respectively. A schematic view of the space partitioning of the solution $\psi^{+} ({\bf r};k)$ is shown in figure~\ref{fig:cluster}.
\hatada{In our notations,
$k=\sqrt{E}$ is the momentum of photoelectron, $E$ is the energy of photoelectron measured from the vacuum level, and $\kappa\equiv\sqrt{k^2-V_0}$ where $V_0$ is a constant potential in the interstitial region.
In this paper, we adopt the condition $\kappa=k=\sqrt{E-V_0}$ without the outer sphere.
}
For $\kappa=k$, the wavefunction $\psi^{+}\, ({\bf r};k)$ satisfies the Lippmann-Schwinger equation 
\begin{equation}
 \psi^{+} ({\bf r};k) = \phi_{0} ({\bf r};k)
 + \int d{\bf r}' \, G^{+}_{0}({\bf r} - {\bf r}' ;k ) \, V({\bf r}') \, \psi^{+} ({\bf r}';k)
\end{equation}
with $\phi_{0} ({\bf r};k)$ the free solution and $V$ ({\bf r}) the potential.
We assume that the wavefunctions
$\psi^{+}_o\, ({\bf r}_o;k)$ and $\psi^{+}_j \, ({\bf r}_j;k)$ respectively outside and inside of the cluster,
can be expanded into local solutions $\Phi_L({\bf r}_o;k)$ and $\Phi_L({\bf r}_j;k)$ as
\begin{eqnarray}
 \psi^{+}_j \, ({\bf r}_j;k)
 &=&
 \sum_L A_{L}^{\,j}({\bf k })\,\Phi_L({\bf r}_j;k)
 \label{eq:psi+_j} \\
 \psi^{+}_o \, ({\bf r}_o;k)
 &=&
 \sum_L \tilde{ A }_{L}^{\,o}({\bf k })\,J_L({\bf r}_o;k)
 +
 \sum_L A_{L}^{\,o}({\bf k })\,\Phi_L({\bf r}_o;k)
\end{eqnarray}
where 
$ \tilde{ A }_{L}^{\,o}({\bf k }) \equiv
 i^{\,l} \, \mathcal{ Y }_L(\hat{\bf k }) \, \sqrt{k/\pi} $.
 \ota{
We employ the notation $L=(l,m)$
for the set of the the angular azimuthal quantum number $l$ and magnetic quantum number $m$. The 
$\mathcal{ Y }_L $ are the real spherical harmonics
defined as a linear combination of the standard complex spherical harmonics $Y_{L}$ \cite{Gianturco1986,Brouder1989}
as
\begin{equation}
\mathcal{Y}_{L}(\hat{\mathbf{k}})
 =\sum_{m^{\prime}} C_{m m^{\prime}}^{l}Y_{lm^{\prime}}(\hat{\mathbf{k}}) 
\end{equation}
where $C_{m m^{\prime}}$ for $m>0$ are
\[ C_{m m}^{l}=(-1)^{m} / \sqrt{2}, \quad C_{m-m}^{l}=1 / \sqrt{2} \]
\[ C_{-m m}^{l}=-i(-1)^{m} / \sqrt{2}, \quad C_{-m-m}^{l}=i / \sqrt{2},\]
and $C_{00}^{l}=1$, otherwise $C_{m m^{\prime}}^{l}=0$.
 }
%
The functions
$J_{L}({\bf r};k)$ and $\tilde {H}_{L}^{+} ({\bf r};k)$
are defined as
$
 J_{L}({\bf r};k)
 \equiv
 j_{l}\,(k r) \,
 \mathcal{ Y }_L (\hat{\bf r})
$
and
$
 \tilde {H}_{L}^{+} ({\bf r};k)
 \equiv
 -i k \,
 h_{l}\,(k r) \,
 \mathcal{ Y }_L (\hat{\bf r})
 \,
$
where $j_{l}$ and $h_{l}$ are respectively the spherical Bessel function and the spherical Hankel function of the first kind of order $l$.
The expansion of the local solution \ota{into real spherical harmonics} is 
\begin{equation}
\Phi_{L}({\bf r}_j;k)
 =
 \sum_{L'} R_{L'L}^{\,j} (r_j;k) \,
 \mathcal{ Y }_{L'} (\hat{\bf r}_j).
\end{equation}
\hatada{In order to obtain the radial solution $R_{L'L}^{\,j}$,
we employ the 3D Numerov method~\cite{Hatada2010} to solve the three dimensional local Schr\"odinger equation. This avoids expanding the truncated potential into spherical harmonics, which is known to lead to Gibbs-type~\cite{Williams1974} convergence problems. 
}
\par
\ota{
The matrix elements of the free electron Green's function in equation~\ref{eq:equality} expanded into real spherical harmonics on each site are called the real-space KKR structure factors and are given by~\cite{Natoli2003}
\begin{equation}
  G_{LL'}^{ij} \,
 \equiv
 4 \pi
 \sum_{L''}
 i^{\,l-l'+l''}
  C(L, L', L'') \, 
 \tilde{H}_{L''}^{+} ({\bf R}_{ij};k),
 \label{eq:GLL}
\end{equation}
where $\otarev{C(L, L', L'' )} \equiv \int d\hat{\bf r} \mathcal{ Y }_{L} (\hat{\bf r})\mathcal{ Y }_{L'} (\hat{\bf r})\mathcal{ Y }_{L"} (\hat{\bf r}) $ is a Gaunt coefficient of the real spherical harmonics. }
\par
\ota{
On the basis of Green's theorem, the sum of the surface
integrals on the truncated cells of the cluster in equation~\ref{eq:equality} is equal to the surface
integral on its bounding sphere (the equality is valid also if $V_0 \ne 0$ as shown 
in Ref.~\cite{Hatada2010}).}

Equation~\ref{eq:equality} can then be reformulated in terms of the matrices $S$
and $E$ (which asymptotically behave as the sine and exponential function respectively)~\cite{Gonis2000}
\begin{eqnarray}
 S_{L'L}^i
 &\equiv
 (R_b^i)^2 \,
 \mathcal{ W }
 \left[
 j_{l'} (k r_i) \,
 ,
 R_{L'L}^{\,i} (r_i;k) \,
 \right]_{r_i=R_b^i},
  \\
 E_{L'L}^{\,i}
 &\equiv
 (R_b^i)^2 \,
 \mathcal{ W }
 \left[
 -ik h_{l'} (k r_i) \,
 ,
 R_{L'L}^{\,i} (r_i;k) \,
 \right]_{r_i=R_b^i}.
\end{eqnarray}
\ota{Here, ${\mathcal{ W }}$ is the Wronskian at the surface of the bounding sphere of scattering site $i$, and $R_b^i$ is the corresponding radius (See figure~\ref{fig:cluster})}.

\ota{
At this stage, we introduce the transition operator (or scattering $T$ matrix) defined by}
\begin{equation}
 T^{\,i}
\equiv
 -S^{\,i}\,(E^{\,i})^{-1} \label{eq:Tmatrix} 
\end{equation}
\ota{
and the coefficients $B_{L}^{i}$ 
related to the coefficients $A_{L}^{i}$ by
\begin{equation}
B_{L}^{i}(\mathbf{k}) \equiv \sum_{L^{\prime}} S_{L L^{\prime}}^{i} A_{L^{\prime}}^{i}(\mathbf{k}). \label{eq:coefficientB}
\end{equation}
}
%
Inserting equations~\ref{eq:psi+_j} and \ref{eq:GLL} in equality equation~\ref{eq:equality}
\ota{and using equations~\ref{eq:Tmatrix} and \ref{eq:coefficientB}}, we obtain the following expression for the wavefunction,
\begin{equation}
 \psi^{+}_{i}({\bf r}_i;k)
 =
 \sum_{L}
 B_{L}^{\,i} ({\bf k})
 \bar{\Phi}_L ({\bf r}_i)
 \label{eq:Psi+}
\end{equation}
where
\begin{eqnarray}
 B_{L}^{\,i} ({\bf k})
 &=&
 \sum_{jL'}
 \tau_{LL'}^{ij} \,
 I_{L'}^{\,j} ({\bf k}), \\
\bar{\Phi}_L ({\bf r}_i)
 &=&
 \sum_{L'} (\,^t{S})_{LL'}^{-1} \Phi_{L'} ({\bf r}_i), \\
 \tau &\equiv &
 \left(
 T^{-1}
 -
 G
 \right)^{-1}, \label{eq:tau}
 \\
 I_{L}^{\,i} ({\bf k})  
 &\equiv& 
 i^{\,l}
 \sqrt{\frac{k}{\pi}}\, 
 e^{i{\bf k }\cdot{\bf R}_{io}}
 \mathcal{ Y }_{L}(\hat{\bf k })\hatadarev{.}
\end{eqnarray}
 Here, $^tS$ denotes the transpose of the matrix $S$.
\hatada{The amplitudes $B_{L}^{\,i} ({\bf k})$ are determined by the constraint that the wavefunction and its first derivative match smoothly to the outer multiple scattering scheme at the surface at each cell. The multiple scattering matrix $\tau$ (or scattering path operator) is constructed from the transition operator and free electron Green's function defined in equation~\ref{eq:GLL}.}

\subsection{Expression of Photoelectron Diffraction Cross-section}
In this subsection, we study the photoelectron diffraction cross-section based on Fermi's golden rule \hatadarev{within the single channel approximation}.
We assume that the wavefunctions of the core electron, photoelectron, and valence electrons
do not mix-up and can be treated independently\hatadarev{.}
Under the conditions mentioned above
and within the electric dipole approximation,
\hatadarev{the many body electric dipole matrix element reduces to a single particle one. Then}
the general expression of the photoelectron diffraction cross-section
along direction $\hat{\bf k}$ from an emitter atom on site $i$, 
is given by
\begin{equation}
 I({\bf k},\hat{\bf \epsilon})
 =\,
 8 \pi^2 \alpha \hbar \omega
 \sum_{m_c}
\left| \,
 \braket{ \,
 \hatadarev{\psi^{-}_i} ({\bf r}_i;{\bf k}) | \,
 \hat{\bf \varepsilon} \cdot {\bf r}_i \, | \,
 \phi_{L_c}^{\,c} ({\bf r}_i)
 }
 \right|^2,
\end{equation}
where \hatadarev{$\phi_{L_c}^{\,c} ({\bf r})$ with the short-hand notation $L_c=(l_c,m_c)$ is the initial core state,}
$\alpha$ is the fine structure constant, $\hbar \omega$ is the photon energy, $\hat{\bf \varepsilon}$ is the polarisation vector of photon.
We neglect the spin polarisation in the photoemission process, therefore introducing a factor two into the formula above.

Using the multiple scattering expression for the photoelectron wavefunction $\hatadarev{\psi^{+}_i} ({\bf r}_i;{\bf k})$ in equation~\ref{eq:Psi+}, we obtain the following formula for the MFPADs
\begin{equation}
 I({\bf k},\hat{\bf \varepsilon}) 
  =
 8 \pi^2 \alpha \hbar \omega
 \sum_{m_c}
\left| \,
 \displaystyle{
 \sum_{L}
 B_L^{\,i*} ({\bf k}) \,
 M_{L_cL} (\hat{\bf \varepsilon})
 }
 \right|^2,
\end{equation}
where
\begin{eqnarray}
 M_{L_cL} (\hat{\bf \varepsilon})
 &\equiv&
 \int d{\bf r}
 \, \bar{\Phi}_L({\bf r}) \,
 \hat{\bf \varepsilon} \cdot {\bf r} \,
 \phi_{L_c}^{\,c} ({\bf r}), \, \nonumber
\end{eqnarray}
is the transition matrix which describes the excitation by 
X-rays of polarization $\hat{\bf \varepsilon}$ of an electron from a core orbital $\phi_{L_c}^{\,c}$.

\subsection{Polarization Average of Photoelectron Cross-section}
The polarization average of the MFPADs is obtained by angular integration of the polarization
\begin{eqnarray*}
 \left< I ({\bf k})\right>_{\varepsilon} 
 &\equiv &
 \frac{1}{4\pi} \int d\hat{\varepsilon} \, 
 I({\bf k},\hat{\varepsilon})
 \\
 &=&
 2 \pi \alpha \hbar \omega
 \sum_{m_cLL'}
 B_L^{\,i*} ({\bf k})\,
 B_{L'}^{\,i} ({\bf k})\, 
 \int d\hat{\varepsilon} \, 
 M_{L_cL} (\hat{\varepsilon}) \,
 M_{L_cL'}^{*} (\hat{\varepsilon}).
\end{eqnarray*}
Since the electric dipole operator can be expanded as $ \hat{\bf \varepsilon} \cdot {\bf r}=(4\pi/3) \, r \sum_m \mathcal{ Y }_{1m}(\hat{\bf \varepsilon}) \mathcal{ Y }_{1m}( {\hat{\bf r}})$~\cite{Varshalovich1988}, the integration reduces to a sum of Cartesian components,
\begin{eqnarray*}
&& \int d \hat{\varepsilon} \, 
 M_{L_cL} (\hat{\varepsilon}) \,
 M_{L_cL'}^{*} (\hat{\varepsilon})\nonumber \\
 &=& 
 \frac{4\pi}{3} 
 \left[
 M_{L_cL} ({\hat{\bf x}})
 M_{L_cL'}^{*} ({\hat{\bf x}}) 
 +
 M_{L_cL} ({\hat{\bf y}})
 M_{L_cL'}^{*} ({\hat{\bf y}}) 
 +
 M_{L_cL} ({\hat{\bf z}}) 
 M_{L_cL'}^{*} ({\hat{\bf z}}) 
 \right]. \nonumber
\end{eqnarray*}
Thus, we obtain the following expression for the PA-MFPADs 
\begin{eqnarray}
 \left< I ({\bf k})\right>_{\varepsilon} 
 &=&
 \frac{1}{3}
 \left[
 I({\bf k},\hat{\bf x})
 +
 I({\bf k},\hat{\bf y})
 +
 I({\bf k},\hat{\bf z})
\right]. \nonumber 
\end{eqnarray}
We see that averaged intensity of the MFPADs over the polarization vector along $x$, $y$ and $z$-axes reduces to the intensity of the PA-MFPADs.

The intensity of the the PA-MFPADs denoted as
$ \left< I ({\bf k})\right>_{\varepsilon} $ is given as
\begin{eqnarray}
 \left< I ({\bf k})\right>_{\varepsilon} 
 &=&
 \frac{8 \pi^2 \alpha \hbar \omega}{3}
 \sum_{n,m_c}  
 \Bigg|
 \sqrt{
 \frac{4\pi}{3}
 }
 \sum_{LL'}
 B_L^{\,i*} ({\bf k})
 \nonumber \\
 &&
 \hspace{0.3cm}
 \times
 \otarev{C(L_c, 1n, L')} \int dr \,
 r^3 \,
 R_{L'L} (r;k) \,
 R_{L_c}^{\,c} (r) \,
 \,
\Bigg|^2.
\end{eqnarray}
\begin{figure}[ht]
\includegraphics[width=0.75\linewidth]{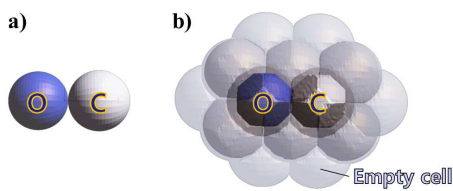}
\caption{
\label{fig:cells}
Cell images of CO used for our multiple scattering calculations,
 (a) for the Muffin-tin approximation with two spherical non overlapping atomic spheres and
 (b) for the Full-potential method with two Voronoi atomic cells, and 22 truncated spherical cells (the so-called empty cells) which contain only electron charge density (no nucleus inside).
The positions of cells are arranged in a BCC structure.
}
\end{figure}
\begin{figure*}[ht]
\includegraphics[width=0.7\linewidth]{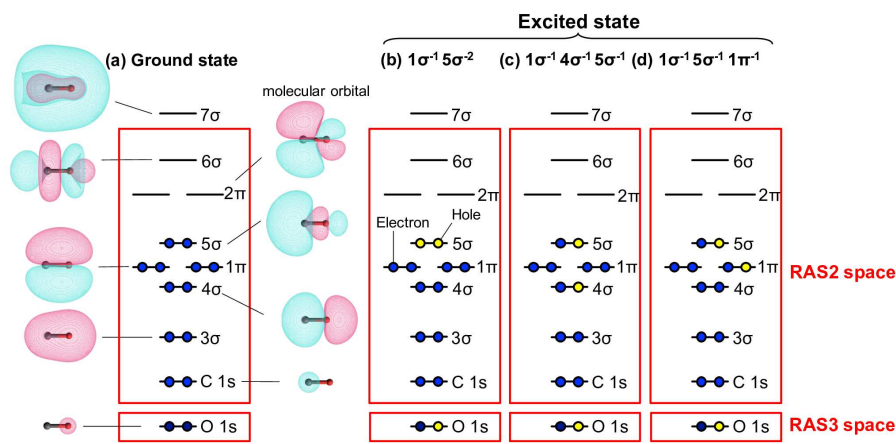}
\caption{
\label{fig:electron_config} 
Major electronic configurations of 
(a) the ground state and the 
 most probable excited states
 after oxygen $1s$ excitation
(b) 1$\sigma^{-1} 5\sigma^{-2}$ ,
(c) 1$\sigma^{-1} 4\sigma^{-1} 5\sigma^{-1}$ and
(d) 1$\sigma^{-1} 5\sigma^{-1} 1\pi^{-1}$, respectively.
The excited states (b), (c) and (d) are 
\hatadarev{the dicationic states populated via the C ($1s$) KVV Auger decay with a core hole created by the second pulse in the oxygen 1$s$ orbital, that is to say CO$^{3+}$.}
Blue and yellow dots in the energy levels represent respectively electrons and holes. 
The shapes of the molecular orbitals are illustrated.
Active spaces (RAS2 and RAS3 space) for RASSCF calculations are assigned in the red boxes.
}
\end{figure*}

\section{Computational Details}
\label{sec:Comp}
\subsection{Quantum Chemistry Calculation of the Electronic Structure}
In the following, we specialize to a two-color XFEL pump-probe experiment. 
The pump pulse removes an electron from the 2$\sigma$ (C($1s$)) orbital of
CO which results in the subsequent Auger ionization to, e.g., the singlet
5$\sigma^{-2}$ states~\cite{Kelber1981,Feyer2005,Cederbaum1991}.
The probe pulse further eliminates an electron from the 1$\sigma^{-1}$ (O($1s$)) orbital. This pump-probe process finally produces a doublet 1$\sigma^{-1}$5$\sigma^{-2}$ state.
All electronic structure calculations in CO were carried out with the MOLCAS 8.2 quantum chemical package~\cite{Aquilante2016}.
The electrostatic potential of the doublet 1$\sigma^{-1}\sigma^{-2}$ state of CO$^{3+*}$ as a function of C-O bond length for the multiple scattering calculation were constructed at the multireference second-order perturbation theory with restricted active space (RASPT2) level of theory~\cite{Malmqvist2008,Sauri2011}.
A ionization potential electron affinity (IPEA) shift of 0.25 a.u. was introduced to reduce the intruder problem~\cite{Ghigo2004}.
The reference wavefunctions for the RASPT2 calculations were computed at the state-averaged Restricted Active Space self-consistent field (SA-RASSCF) level of theory~\cite{Malmqvist1990}.
The 10-low-lying doublet states were considered without symmetry, and their average energy was minimized with the equal weight. For all RASPT2 and SA-RASSCF calculations, we included 10 spatial orbitals constructed mainly from $1s$, $2s$, $2p$ orbitals of carbon and oxygen atoms, that is 1$\sigma$, 2$\sigma$, 3$\sigma$, 4$\sigma$, 2 $\times$ 1$\pi$, 5$\sigma$ (HOMO), 2 $\times$ 2$\pi$ (LUMO) and 6$\sigma$ as active space, where the maximum occupation of the 1$\sigma$ orbital is restricted to one to describe the core-hole in this orbital. The 11 electrons in the active orbitals were treated as active electrons. The ANO-RCC-VQZP basis set ~\cite{Roos2004} was consistently used. Relativistic effects were taken into account through the second order Douglas-Kroll-Hess~\cite{Peng2012}
and mean-field spin-orbit~\cite{Hess1996} Hamiltonians.\\

This RASPT2 treatment is essential for an accurate evaluation of the chemical shift in photoelectron energy and photoelectron momentum. The second order perturbation and relativistic corrections affect only the energetics in this system. The core orbital is identical to the O(1s) Hartree-Fock (HF) orbital of CO$^{2+}$.  We used the natural orbitals obtained from RASPT2 calculations for the input of the electron density calculations, however the RASPT2 natural orbitals are almost identical to those from RASSCF. Thus, the RASPT2 density is almost identical to the RASSCF one, and the electron density triple cation is close to that of a HF calculation on the ion especially for the 1$\sigma^{-1}$ 5$\sigma^{-2}$ state in the Franck-Condon region. We also calculated the 1$\sigma^{-1}$4$\sigma^{-1}$5$\sigma^{-1}$ and 1$\sigma^{-1}$5$\sigma^{-1}$1$\pi^{-1}$ states for comparison. However, they have multiconfiguration characters and RASSCF (or RASPT2) density is essentially different from HF ones. In addition, the contribution of the 1$\sigma^{-1}$ 5$\sigma^{-2}$ configuration on the RASSCF wavefunction varies when we elongate 
the C-O bond length. This is another reason that the RASSCF (or RASPT2) treatment is essential.

\begin{figure*}[ht]
\includegraphics[width=0.8\linewidth]{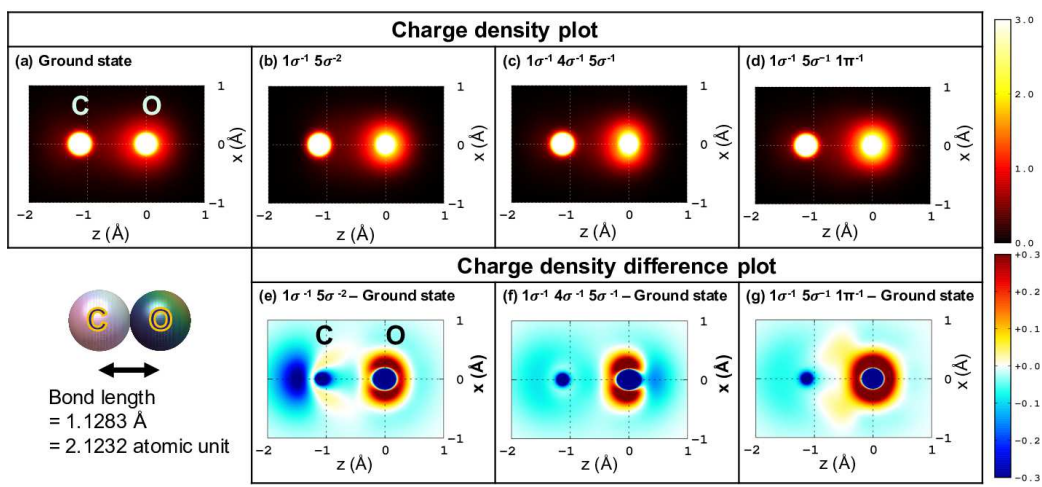}
\caption{
\label{fig:dens_diff-dens} 
Electron charge density plots of CO
in \ota{the molecular axis} plane 
 for
 (a) the ground state
 and the excited states
 (b) $1\sigma^{-1} 5\sigma^{-2}$,
 (c) $1\sigma^{-1} 4\sigma^{-1} 5\sigma^{-1}$
 and
 (d) $1\sigma^{-1} 5\sigma^{-1} 1\pi^{-1}$
 defined in figure~\ref {fig:electron_config}.
 \hatadarev{These are the dicationic states populated via the C ($1s$) KVV Auger decay with a core hole created by the second pulse in the oxygen 1$s$ orbital, that is to say CO$^{3+}$.}
The electronic structures are calculated
with the ANO-RCC-VQZP basis set,
where the oxygen atom is set at the origin and the C-O bond length is fixed to 1.1283 \mbox{\AA} for all the calculations.
the oxygen $1s$ hole in the RAS3 space is kept during the calculations for the excited states.
The plots (e)-(g) are electron charge density difference plots between (a) the ground state and the corresponding excited state among (b)-(d), respectively.
}
\end{figure*}
\begin{figure}[ht]
\begin{center}
\includegraphics[width=0.9\linewidth]{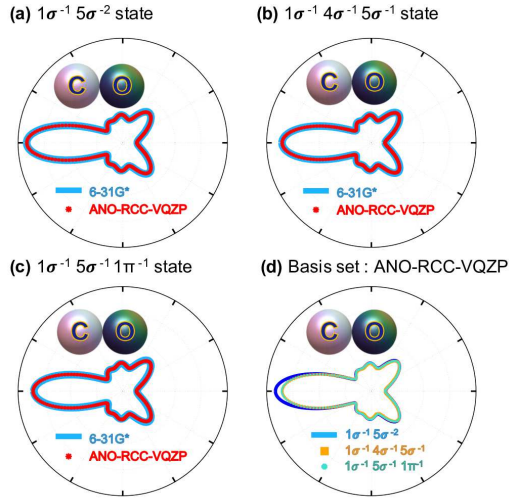}
 \caption{
\label{fig:basis_config} 
Influence of the basis functions on the PA-MFPADs at 100 {\rm eV}
kinetic energy of the photoelectron with 1.1283 \mbox{\AA} of C-O bond length (equilibrium bond length in ground state).
\hatada{The calculations are performed within the Full-potential method using 6-31${\rm G^{*}}$ and the ANO-RCC-VQZP basis sets}.
(a) $1\sigma^{-1}$\,$5\sigma^{-2}$, (b) $1\sigma^{-1}$\,$4\sigma^{-1}$\,$5\sigma^{-1}$ and 
(c) $1\sigma^{-1}$\,$5\sigma^{-1}$\,$1\pi^{-1}$ state respectively.
 \hatadarev{These are the dicationic states populated via the C ($1s$) KVV Auger decay with a core hole created by the second pulse in the oxygen 1$s$ orbital, that is to say CO$^{3+}$.}
(d) Influence of the electronic structure type on the PA-MFPADs.
\otarev{
We compare the PA-MFPADs for the following electronic structures:
$1\sigma^{-1}$\,$5\sigma^{-2}$, $1\sigma^{-1}$\,$4\sigma^{-1}$\,$5\sigma^{-1}$ and 
$1\sigma^{-1}$\,$5\sigma^{-1}$\,$1\pi^{-1}$ state where the basis set is 
ANO-RCC-VQZP.
}
}
\end{center}
\end{figure}

\begin{figure*}[ht]
\includegraphics[width=0.7\linewidth]{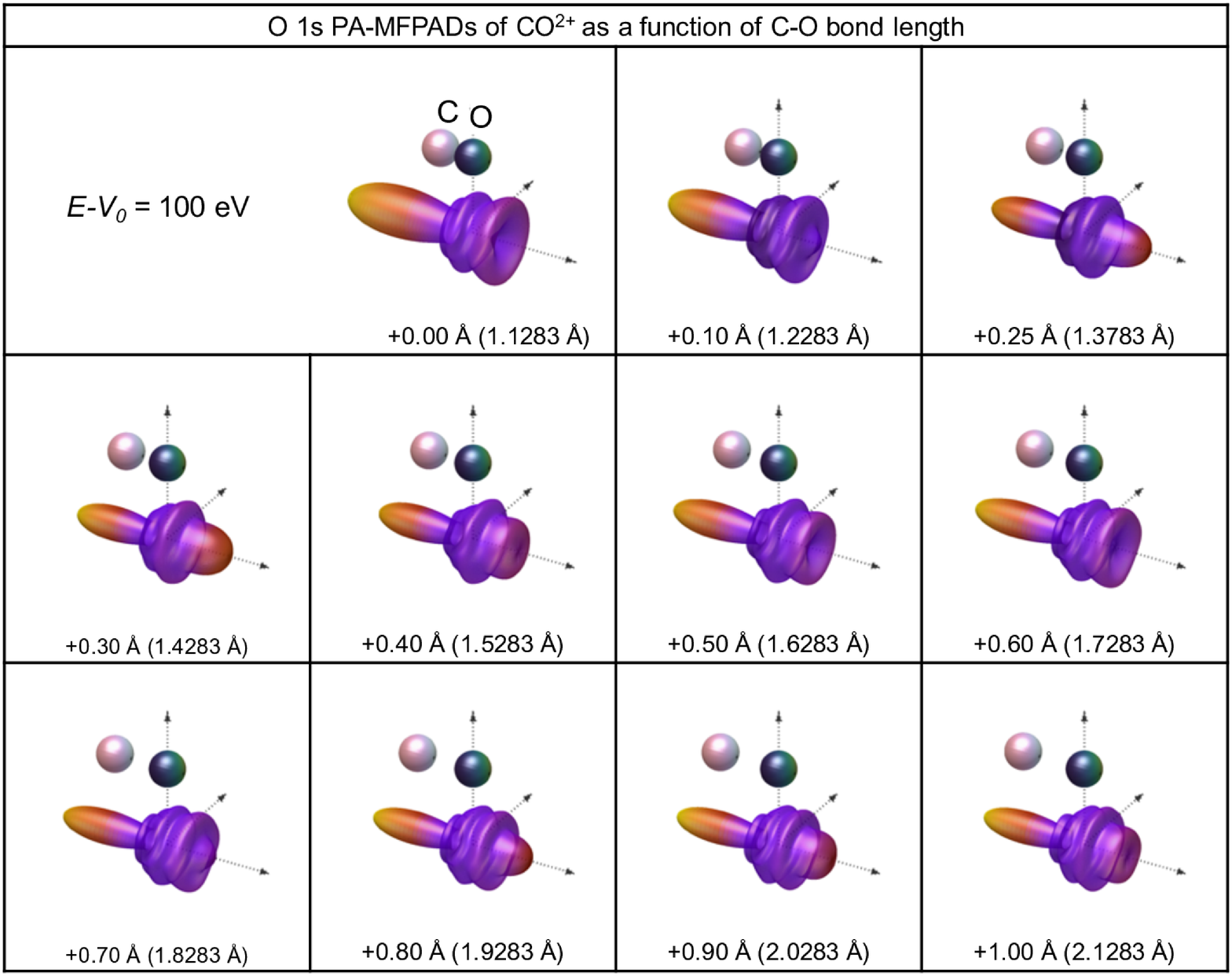}
\caption{
\label{fig:bondlength_dependency} 
The oxygen $1s$ PA-MFPADs of CO$^{2+}$ as a function of the C-O bond length $R$ from 1.1283 \AA\ (equilibrium bond length in ground state) to 2.1283 \mbox{\AA}.
All the calculations were performed within the Full-potential method for the excited state 1$\sigma^{-1}$ 5$\sigma^{-2}$ (i.e. O $1s^{-1}$ HOMO$^{-2}$ \hatadarev{which is triply charged CO, namely CO$^{3+}$}) with a photoelectron kinetic energy of 100 eV. 
}
\end{figure*}

\subsection{Multiple Scattering Calculation for Full-potential and Muffin-tin cases}
\label{sec:MScalc}
The scattering matrix $\tau$ and the transition matrices $T$ were  calculated with the FPMS code~\cite{Hatada2007,Hatada2010}
within the framework of the multiple scattering approach.
The photoelectron diffraction calculations were performed with the MsSpec code~\cite{Sebilleau2011}, using these matrices as an input.
For Full-potential calculations, we need to add empty cells (pseudo atomic cells with electron charge density but no nucleus) 
in order to account for the charge density outside the (Voronoi) atomic cells.
We have used 22 empty cells to embed the carbon and oxygen cells within a BCC structure as shown in figure~\ref{fig:cells}. \hatada{The multiple scattering calculation was found to converge at $l_{max}=6$ and 8 for the Muffin-tin approximation and the Full-potential method, respectively. The same $l_{max}$ was used for the expansion of the wavefunction $\psi^+_j ({\bf r};k)$ into local solutions and the expansion of each local solution into spherical harmonics.}
%
\hatada{
Following Kazama {\it et al.}~\cite{Kazama2012}, the kinetic energy of the photoelectron was estimated as $k=\sqrt{E-V_0}$, where $V_0$ is the surface average of the potential of the cluster.} \hatadarev{This $V_0$ is the threshold for the continuum and bound state.}
\hatadarev{The wavefunction outside the cells is a plane wave of momentum $k$.}

\hatadarev{In order to estimate the $T$-matrices for each cell, we solved the local Schr\"odinger equations with an optical potential which is the sum of the static Coulomb potential, namely the Hartree potential, and the exchange-correlation potential.}
To take into account the electron correlation between the ejected core electron and the $N-1$ remaining electrons, we used a Hedin-Lundqvist potential~\cite{Hedin1969,Hedin1971} \hatadarev{which is a self-energy used in the field of solid state physics}. This potential is  \hatada{based on} the $GW$ approximation with a non-SCF \hatada{Green's function $G$, a screened potential $W$} and a single plasmon pole approximation, and allows to model satisfactorily the scattering optical potential in the multiple scattering calculations.
It is known however that its imaginary part produces a rather strong damping effect~\cite{Hatada2010,Hatada2007}, especially for light elements such  \hatada{as carbon and oxygen}, hence we used only the real part in our calculations.
Since CO molecules do not have any particular direction perpendicular to the molecular axis,
the intensity of photoelectron was averaged over the azimuth angle around that axis.

%
\par
The electron charge density for the multiple scattering calculations has been prepared following two different prescriptions.
In the first one, which we refer to as ''non-SCF'' (non-self-consistent), the molecular potential and the electron charge density were constructed by a simple superposition of the free atomic densities. The free atom calculations were performed with the Desclaux code~\cite{Desclaux1975} which employs a multi-configuration relativistic Dirac-Fock theory, and the excited states were calculated with a fully relaxed core hole.
The second one, which we call ''SCF'', 
\hatada{employs the electron charge density calculated from MOLCAS 8.2 with the help of a Hermite Gaussian basis functions set~\cite{Helgaker2000} and the (Coulomb) static part of the scattering potential calculated under the McMurchie-Davidson scheme~\cite{Helgaker2000} with a Boys function~\cite{Boys1950}. The details can be found in Ref.~\cite{Komiya2018}.}

In the Muffin-tin calculations, for both prescriptions, the electron charge density and the static part of the potential were spherically averaged. For the Muffin-tin spheres, we used touching but non overlapping C and O atomic spheres. The sphere radii were estimated for each bond length by the Norman criterion~\cite{Norman1974} and  are roughly proportional to the ionic radius of the atom in the periodic table.
An empirical prescription to improve the results is to use overlapping spheres, 10$\sim$15\%, to account approximately for the presence of covalent electrons (see Refs.~\cite{Herman1974,Norman1974}). This approximation is called the Atomic-Sphere Approximation (ASA) in LMTO and KKR~\cite{Turek1996}. However, as it is not our scope in this work to empirically optimise the results through tentative overlapping, we have kept to the touching spheres. \hatada{For $V_0$, we have used the value obtained from our Full-potential calculations.}

\section{Results}
\label{sec:Results}
\begin{figure}[ht]
\includegraphics[width=\linewidth]{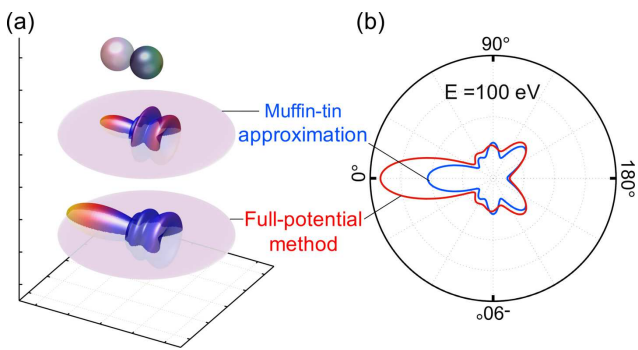}
\caption{
 \label{fig:MTandFP} 
The oxygen $1s$ PA-MFPADs of CO$^{2+}$ in the excited state 
$1\sigma^{-1} 5 \sigma^{-2}$ 
\ota{(i.e. O $1s^{-1}$ HOMO$^{-2}$ \hatadarev{which is triply charged CO, namely CO$^{3+}$})} 
\ota{with $100\,{\rm eV}$ of photoelectron kinetic energy}.
The oxygen atom is located at the origin and the azimuthal angle is defined from the direction \ota{of the carbon atom}.
(a) 
The oxygen $1s$ PA-MFPADs
 within the Muffin-tin approximation is shown in the upper plane and that 
 within the Full-potential method
 is displayed in lower plane.
(b) Comparison of the PA-MFPADs 
between 
Muffin-tin approximation (blue line) and Full-potential method (red line).
 }
\end{figure}
%
\subsection{Dependence on Electronic Structures}

\ueda{We consider three states\hatada{, 1$\sigma^{-1} 5\sigma^{-2}$,
1$\sigma^{-1} 4\sigma^{-1} 5\sigma^{-1}$ and
1$\sigma^{-1} 5\sigma^{-1} 1\pi^{-1}$,}
as shown in figure~\ref{fig:electron_config}.} 
These states are the dicationic states populated via the C \hatadarev{($1s$)} KVV Auger decay ~\cite{Cederbaum1991} with a core hole created by the second pulse in the oxygen 1$s$ orbital.

In figure~\ref{fig:dens_diff-dens} the ground state and excited states electron charge densities are displayed in the upper row and the electron charge density difference plots, between each excited state
and the ground state, are shown in lower row.
\hatada{As 1$\sigma^{-1} 4\sigma^{-1} 5\sigma^{-1}$ in \hatadarev{(c)} has a \hatadarev{$C_{2v}$} symmetry \ota{around} the molecular axis, 
we chose the $xz$-plane to lay the orbital in order to see its anisotropy. All the other electronic configurations have a \hatadarev{$C_{\infty v}$} symmetry \ota{around} the axis. 
These comparisons show the change of charge distributions varying with electronic configuration.}

\hatada{Next, we studied the influence of the electronic structure on the PA-MFPADs.}
Figure~\ref{fig:basis_config} shows the results of PA-MFPADs simulations performed for the three electron configurations
using the 6-31G* \yamazaki{(small, double-zeta + valence polarization)} and the ANO-RCC-VQZP \yamazaki{(large, quadratic-zeta + valence polarization)} basis sets.
The results clearly indicate that the PA-MFPADs are not very sensitive to the electronic structure \hatada{in this high energy regime}.
Therefore, in the following subsections, we will use the $1\sigma^{-1}$\,$5\sigma^{-2}$ configuration as a representative electronic configuration
using the ANO-RCC-VQZP.

\subsection{Dependence on Bond Length}

%
The PA-MFPADs calculated as a function of the  
\ota{C-O bond length}
are shown in figure~\ref{fig:bondlength_dependency}. 
The forward-\hatada{intensity} $\left< I (k,\theta=0^{\circ})\right>_{\varepsilon}$ corresponds to the largest  lobe, which is directed  along the bond towards \ota{the carbon atom}. This lobe is due to the so-called focusing effect of photoelectrons by the carbon atom. 
The backward-\hatada{intensity} $\left< I (k,\theta=180^{\circ})\right>_{\varepsilon}$ is in the opposite direction, and it oscillates between the peak top and the valley of the lobe as a function of $R$.
\ueda {Within the framework of the Plane Wave  single-scattering approximation of the electron Green's function in equation~\ref{eq:GLL} (see for instance the EXAFS formula~\cite{Rehr1986}), one period of the oscillation corresponds to an increase by $\Delta R =\pi/k$ in $R$.
}
There are several small lobes between the forward and backward directions\ota{, which} move from backward to forward as $R$ increases. \ueda{The number of lobes is incremented every one period of oscillation of the backward-intensity peak. 
Labelling angles $\theta_1$ and $\theta_2$ for 
two neighbouring lobes ($\theta_1<\theta_2$),
we have approximately $R=2\pi/k(\cos\theta_1-\cos\theta_2)$.
Thus, the PA-MFPADs do contain the information about the bond length. We will discuss this relation in more detail in the companion paper}~\cite{Ota2020}.

\subsection{Dependence on Shape of Potential}

\begin{figure*}[htb]
\includegraphics[width=0.7\linewidth]{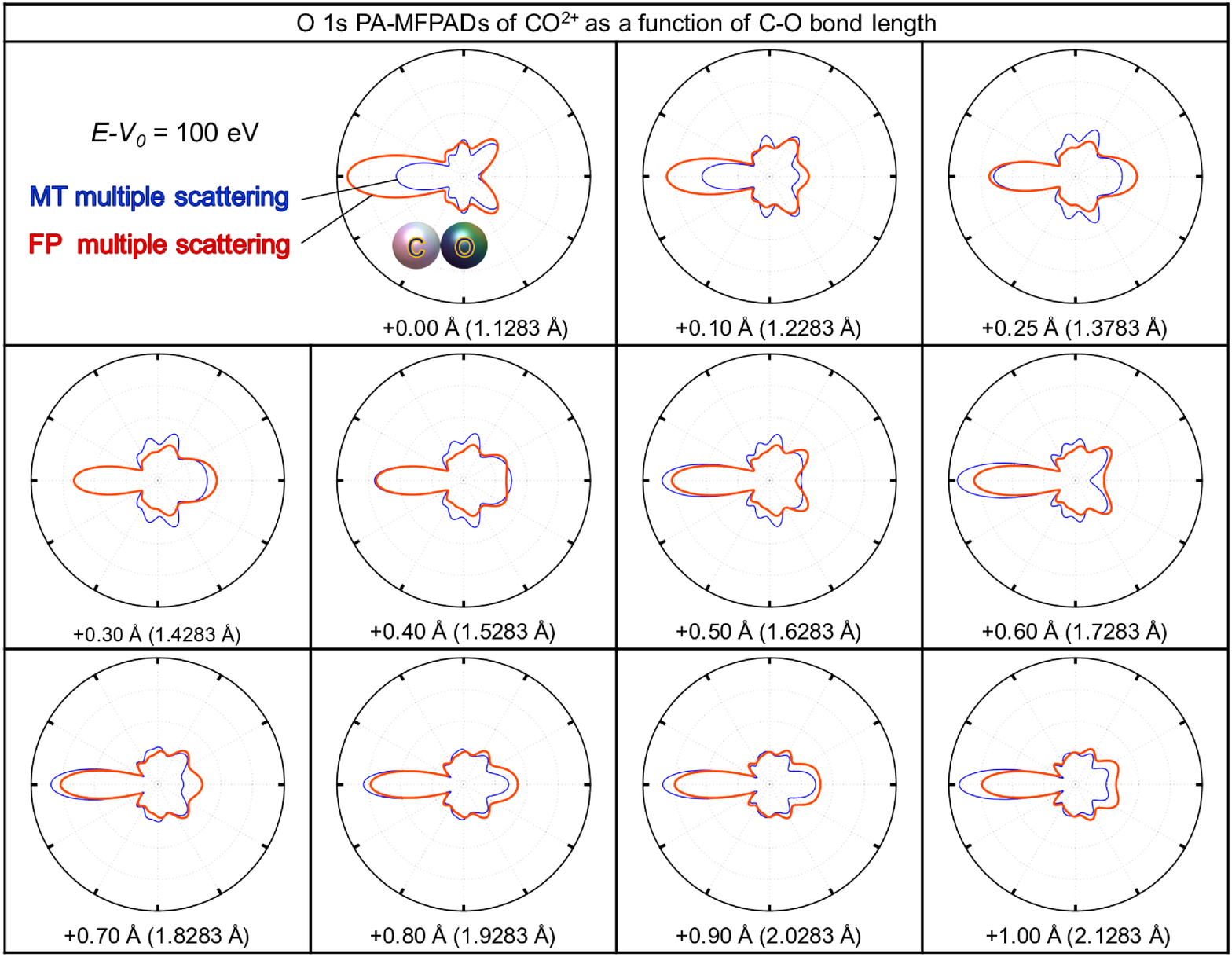}
\caption{
\label{fig:bondlength_MT_FP} 
The oxygen $1s$ PA-MFPADs of CO$^{2+}$ as a function of the C-O bond length $R$ from 1.1283 \AA\ (equilibrium bond length in ground state) to 2.1283 \AA\ for a photoelectron kinetic energy of 100 eV. All the calculations were performed with the excited state $1\sigma^{-1} 5\sigma^{-2}$ (i.e. O $1s^{-1}$ HOMO$^{-2}$ \hatadarev{which is triply charged CO, namely CO$^{3+}$}).
Blue and red lines indicate the results of multiple scattering calculation using the Muffin-tin approximation and the Full-potential method, respectively.
}
\end{figure*}

%
\begin{figure}[htb]
\includegraphics[width=1.0\linewidth]{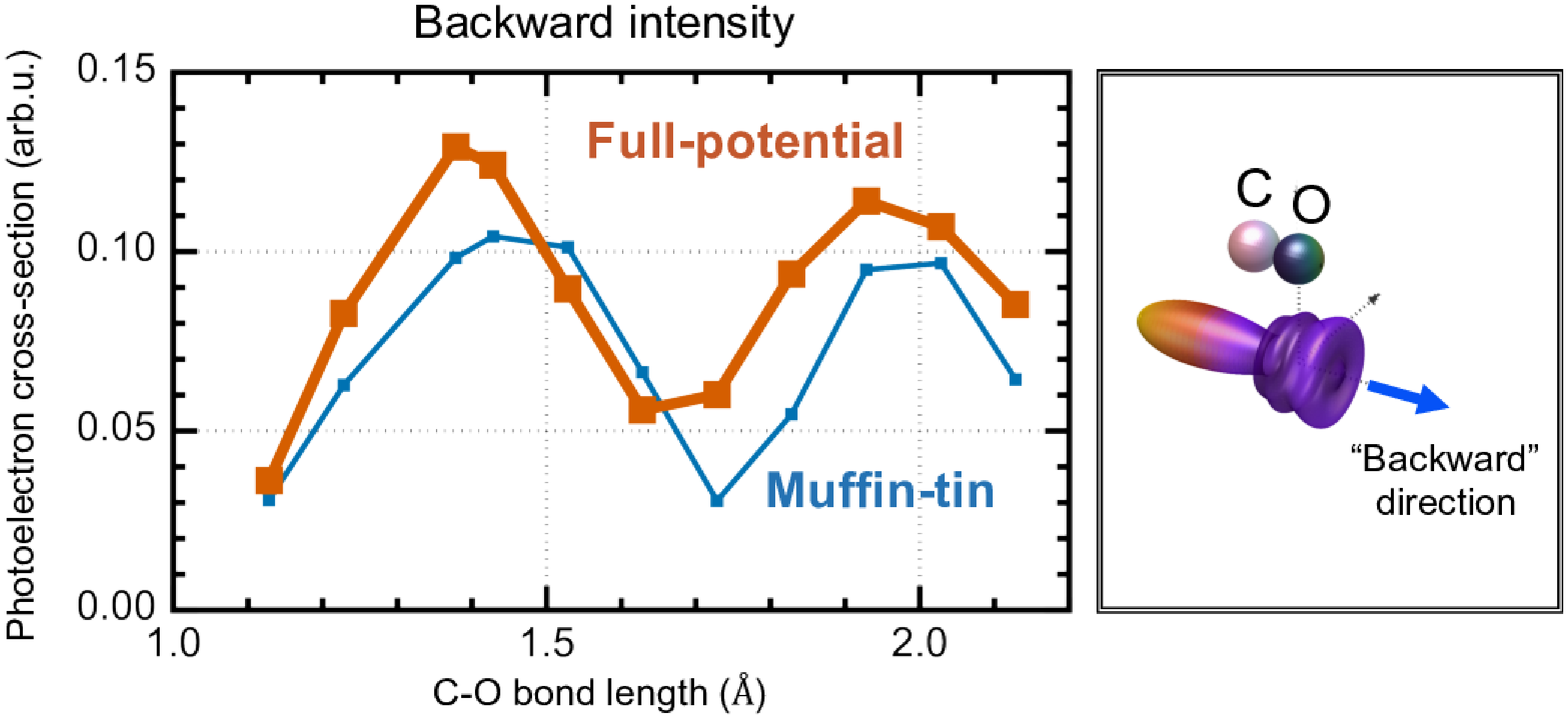}
\caption{
\label{fig:oscillation}
\ota{
Backward-intensities of the oxygen $1s$ PA-MFPADs of CO$^{2+}$ as a function of the C-O bond length $R$ from 1.1283 \AA\, (equilibrium bond length in ground state) to 2.1283 \AA\ for 1$\sigma^{-1} 5\sigma^{-2}$ state  \hatadarev{(i.e. O $1s^{-1}$ HOMO$^{-2}$ which is triply charged CO, namely CO$^{3+}$)}.
Dark blue and orange lines indicate multiple scattering calculation results within the Muffin-tin approximation and the Full-potential method, respectively.
}
}
\end{figure}
In order to evaluate the effect of the Muffin-tin approximation on the PA-MFPADs,
we compare in figure~\ref{fig:MTandFP} the \ota{oxygen $1s$} PA-MFPADs of CO$^{2+}$ calculated  with the Muffin-tin approximation
to the corresponding one obtained from the Full-potential method. 
The major C ($1s$) KVV Auger decay channels in the low-energy region were estimated according to the ADC(2) calculations by Cederbaum {\it et al.}~\cite{{Cederbaum1991}}.
The kinetic energy of the photoelectron is chosen to be 100 eV on the basis of the widely supported idea that photoelectron spectroscopy
is less sensitive to the details of the electronic structure and the electron charge density in a sufficiently high energy regime ($\gtrsim$ 100 eV)~\cite{Hatada2010}.\footnote{\yamazaki{The calculated vertical O($1s^{-1}$) ionization energy from the lowest singlet state of CO$^{2+}$ were found to be 564-569 eV at the RASPT2/ANO-RCC-VQZP level of theory depending the \ota{C-O bond length} $R$ (chemical shifts). This means that we have only 5 \% and 2-3 \% errors on the photoelectron energy and momentum, respectively.}}
We see that the PA-MFPADs patterns are clearly distinct from each other.
The largest contribution to the forward-intensity comes from the forward-scattering \hatada{by the} \ota{carbon atom} \hatada{which induces a focusing effect}.
This forward-scattering is due here to a weak potential. 
\hatada{Since such a weak potential is essentially left outside of the Muffin-tin spheres in the case of short distances,
the forward-intensity is certainly underestimated in the Muffin-tin approximation.}
\hatada{In figure~\ref{fig:bondlength_MT_FP}, we show a series of comparisons between the Muffin-tin approximation and the Full-potential method for multiple scattering calculations of the oxygen $1s$ PA-MFPADs of CO$^{2+}$}
as a function of the C-O bond length $R$ from 1.1283 \AA\ (equilibrium bond length in ground state) to 2.1283 \AA.
All the calculations were performed for the excited state $1\sigma^{-1}\,5\sigma^{-2}$ (i.e. \ueda{O $1s^{-1}$ HOMO$^{-2}$}) with a photoelectron kinetic energy of 100 eV .
\ueda{As noted in figure~\ref{fig:MTandFP}, the forward-intensity peak is significantly underestimated by the Muffin-tin approximation at near equilibrium distance. Besides, the peak positions between the forward- and backward-intensity peaks start to differ with the increase in $R$ between the results in the Muffin-tin approximation and the Full-potential method.}

%
\ota{As we notice in figure~\ref{fig:oscillation}, the frequency of the oscillations of the backward-intensity is almost $2kR$ for both cases. Indeed, this corresponds exactly to the EXAFS oscillation frequency.}
%
%
\ota{
In this high energy regime, i.e. in the EXAFS regime, single scattering towards the backward direction is the main contribution of the backward-intensity, so that we expect that we may observe this EXAFS-like oscillation in the backward-intensity as in EXAFS.
In the case of the oxygen $1s$ PA-MFPADs of CO${}^{2+}$, the back-scattering is caused mainly by the strong (core) potential very near the center of the carbon atom .
At first sight, it may seem strange to have so much difference between the two calculations, since the electron charge densities are almost spherically symmetric in the Full-potential case. In fact, it is the contribution of the forward-scattering by the oxygen atom which accounts for the difference.
This forward-scattering occurs at the second scattering order, along the pathway going from the carbon atom to the detector.
Its influence, especially through the phase of the forward scattering amplitude, $\psi(k,\theta=0^{\circ})$, appears in the EXAFS oscillation as $\cos(2kR + \psi(k,\theta=0^{\circ})+\psi(k,\theta=180^{\circ}))$. \ota{Since} forward-scattering is important even for a weak potential, it is more sensitive to fine details of the electron charge density and therefore, it depends strongly on the modelling of the scattering potential. Thus the error in the Muffin-tin approximation appears to make the difference of peak/valley positions for the two calculations in figure~\ref{fig:oscillation}.
}
%

\subsection{Toward Reaction Dynamics Imaging using PA-MFPADs based on Ab-initio Full-potential Multiple Scattering Theory }

Finally, we would like to consider the potential targets of the reaction dynamics imaging simulations using PA-MFPADs based on the Full-potential multiple scattering theory, just beyond the dissociation of the dication of diatomic molecules.

A possible extension that may be monitored by the present technique is the fragmentation of the methanol dication. There are at least two major fragmentation channels. One is the direct C-O bond breaking~\cite{doi:10.1063/5.0006485}. This process is very similar to the dissociation of the diatomic molecule, which we have considered in the present work.  The other is the molecular dissociation into CO$^+$ and H$_{3}^{+}$. In this process, roaming of H$_2$ takes place and extraction of H$^+$ from CH or OH sites follows ~\cite{Ekanayake2017}. The timescale of this process is $\sim 100$ fs.  We should be able to distinguish between these two pathways, both experimentally and theoretically, by the momenta of fragment ions that define the molecular axis or reaction plane to be observed for O 1s PA-MFPADs via the C(1s) pump O(1s) probe scheme.  

A case a little more complex but still feasible with the present method could be the X-ray induced fragmentation of doubly charged amino acids such as glycine~\cite{PhysRevA.98.053408}, serine~\cite{Itala2016}, etc.  A typical first step of the decomposition is the COOH$^+$ ejection~\cite{Itala2016,Ha2014,doi:10.1063/1.4962061} with a time scale of 10-100 fs~\cite{PhysRevA.98.053408,Itala2016,Ha2014,doi:10.1063/1.4962061}. Capturing the COOH$^+$ loss dynamics could be achieved by the C(1s) pump N(1s) probe scheme.

Furthermore, many photochemical reactions of polyatomic molecules with hetero-atoms for the photoelectron source could also become attractive targets. For example, ultraviolet induced femotosecond H loss of pyrrole via $^1\pi\pi^*$ and $^1\pi\sigma^*$ states \cite{doi:10.1063/1.4742344} will be a suitable target since its N atom is applicable for the photoelectron source and the dissociating H atom locates just next to the N atom. These molecules will also become good potential targets of the imaging experiments with the PA-MFPADs with the two-color XFEL light source.

From the theoretical points of view, the on-the-fly reaction dynamics calculations~\cite{doi:10.1063/1.4896656,doi:10.1063/1.5115072,C6FD00085A,PhysRevA.91.043417,doi:10.1063/5.0006485,Inhester2018,Rudenko2017,Richter2011JCTC,Mai2018WCMS,doi:10.1002/wcms.1158} and ab-initio core-hole spectroscopy \cite{Inhester2018,Hua2019,doi:10.1063/1.3651082,MITANI2003103,doi:10.1063/4.0000016,doi:10.1002/anie.202007192,doi:10.1002/jcc.26219} for these polyatomic molecules are already feasible. 
It will also be helpful to improve the efficiency of the electrostatic potential evaluation and the Full-potential multiple scattering calculations by further tuning including massive parallelization to push its interplay with the experiment and reaction dynamics theory forward. In addition, analytical formulae that extract structural dynamics from the complex scattering results will also be useful~\cite{Ota2020}.

\section{Conclusions}
We have studied theoretically the 
 PA-MFPADs emitted from the $1s$ orbital of oxygen atom of dissociating dicationic carbon monoxide \hatada{CO$^{2+}$}.
\hatada{
Within the framework of multiple scattering theory, we have investigated the influence of the Muffin-tin approximation and the role of the potential in different excited states, with electron charge densities calculated using the RASPT2} 
method in order to model properly the influence of the core hole in the final state.
The Full-potential theory takes into account the electron charge density outside the physical atomic cells by using simultaneously Voronoi polyhedra and truncated spheres.

Despite the fact that the Muffin-tin approximation was giving 
similar results as the Full-potential calculation above 100 eV for XANES~\cite{Hatada2010}, we observe here significant differences in the PA-MFPADs.
This may be explained by the general optical theorem which states for \textit{real potentials} that the integrated photoemission over the whole solid angle equals the X-ray absorption~\cite{Hatada2010,Natoli1986}.
A consequence of this theorem is that in XANES, even if there is a difference in angular distribution between the Muffin-tin approximation and the Full-potential method, this difference will be smeared out by the angular integration.
Moreover, even in the case of long bond lengths, where the atomic approximation works, we found a significant difference in the PA-MFPADs between the Muffin-tin approximation and the Full-potential method.
The consequence is that we need to use the Full-potential multiple scattering theory to study structure dynamics using FELs and Synchrotron even in the high energy region.

We have also studied the PA-MFPADs from different excited states of the Auger final state. We have found that 
the type of excited states does not affect significantly the PA-MFPADs, though the difference in the electronic structure was not negligible.
%
The relationship between the PA-MFPADs patterns and
the C-O bond length
will be discussed in detail
within the multiple scattering theory in the companion  article~\cite{Ota2020}.
%
\section*{Acknowledgement}
\hatadarev{We would like to acknowledge C. R. Natoli for fruitful discussions.}
This work was accomplished under the program of the Dynamic Alliance for Open Innovation Bridging Human, Environment and Materials program and Cooperative Research Program of ``Network Joint Research Center for Materials and Devices''.  K. H. acknowledges funding by JSPS KAKENHI under Grant No. 18K05027 and 17K04980. K. Y. is grateful for the financial support from Building of Consortia for the Development of Human Resources in Science and Technology, MEXT. 

\section*{References}
\bibliographystyle{unsrt}
\bibliography{fukiko}
\end{document}